\documentstyle[aps]{revtex}
\begin{document}
\draft
\def\lsim{\lower.5ex\hbox{$\; \buildrel < \over \sim \;$}}
\def\gsim{\lower.5ex\hbox{$\; \buildrel > \over \sim \;$}}
%\setcounter{page}
%\begin{center}
%\begin{bf}
\title{Difference in the number of operators between coupled and uncoupled
basis for the general SU(n) Lie algebra}
%\vskip0.5cm
%\vskip0.35cm
%\end{large}
%\end{bf}
%\end{center}
%\noindent\\
\author  {Banibrata Mukhopadhyay\\}
\address{S.N. Bose National Centre for Basic Sciences,
Block - JD, Sector - III, Salt Lake, Calcutta - 700091, India\\}

\author{Subhadip Raychaudhuri\\}
\address{Department of Physics and Astronomy, University of Rochester, Rochester
NY 14627, USA\\}
%\end{bf}
\vskip0.5cm
\noindent\\
%\address{Department of Physics, University of Calcutta, 92, A.P.C. Road,
%Calcutta - 700009, India }\\ 	
%e-mail: bm@boson.bose.res.in  \& chakraba@boson.bose.res.in\\
%\thanks{Presently in S.N. Bose National Centre for Basic Sciences,
%Block - JD, Sector - III, Salt Lake, Calcutta - 700091, India\\}
%\thanks{Presently in Department of Physics, University of Rochester, Rochester,
%USA\\}
\thanks{e-mail:  bm@boson.bose.res.in  \& subha@pas.rochester.edu\\}
%\end{center}
%\baselineskip = 12 true pt
\maketitle
\vskip0.3cm
\setcounter{page}{1}
\noindent{Submitted to appear Journal Of Physics A}
\def\ch{\lower-0.55ex\hbox{--}\kern-0.55em{\lower0.15ex\hbox{$h$}}}
\def\lh{\lower-0.55ex\hbox{--}\kern-0.55em{\lower0.15ex\hbox{$\lambda$}}}	
%\begin{centre}
%{\bf Abstract}
%\end{center}
%\vskip0.3cm
%\noindent
%{\it Received ...., accepted ........}\\
%\vskip0.3cm
%\noindent

\begin{abstract}

For the cases of irreducible representation, the complete set of operators
necessary to specify uniquely the states. There are two ways of
representing the state, using uncoupled and coupled basis. Here we discuss,
how the number of operators for the cases of coupled and uncoupled basis
changes as well as their difference with the increase of dimension. For higher
dimensional groups this number difference changes systematically.
 
\end{abstract}


We are familiar with angular momentum in Quantum Mechanics which is 
nothing but $SU(2)$ Lie algebra. When we add two angular momenta $J_1$, $J_2$
to get the total angular momentum $J$ ($= J_1 + J_2$) 
the state of a system can be
represented by two different sets of quantum numbers. One way of representing
the state is by taking the product of individual angular momentum states
and the complete set of quantum numbers are $J_1^2$, $J_{1z}$, $J_2^2$, 
$J_{2z}$. Then we can consider the coupled basis where the quantum numbers are
$J^2$, $J_1^2$, $J_2^2$, $J_z$ which is also a complete set. Now if we
generalise this addition of angular momentum to higher $SU(n)$ groups we can
formally represent as 
$$
D(p_1, q_1)\otimes D(p_2, q_2) = \Sigma \oplus \sigma (p, q) D(p, q)
\eqno{(1)}
$$
where, $D(p_1, q_1)$ and $D(p_2, q_2)$ are two irreducible representation (IR), 
$\sigma(p, q)$ is an integer and corresponding $p_1, q_1, p_2, q_2, p,
q$ are associated with dimensionality of the respective representations.

For $SU(3)$ the complete set of operators necessary to specify uniquely
the states of an IR are $G^3, F^2, I^2 , I_3 , Y$ . So the states
of product representation $D(p_1,q_1) \otimes D(p_2,q_2)$ of two IR
can be completely specified by the eigenvalues of the $10$ linearly
independent operators,

$G^3(1) , G^3(2), F^2(1), F^2(2), I^2(1), I^2(2), I_3(1), I_3(2), Y(1), Y(2)$.

If we define the operators of coupled basis (keeping in mind 
the addition of angular momentum) as 
$$
O_i = O_i(1) + O_i(2) 
\eqno{(2)}
$$
then we get a set of linearly independent commuting operators

$G^3, G^3(1), G^3(2), F^2, F^2(1), F^2(2), I^2, I_3, Y$.

The total number is $9$. So this is a non-complete set. 
We need another operator to make the set complete. Usually
the parity operator is taken to make the set complete.
Now we can consider higher $SU(n)$ groups and try to see the differences
in number of operators when we go from the product basis to coupled
basis. To do that we consider the general $SU(n)$ group.
The general rule for getting number of operators in a complete
set in $SU(n)$ representation is:

there will be $(n-1)$ number Casimir operators (since rank is $n-1$),
$n-1$ number weight and number of Casimir operator for each lower group
of $SU(n)$ i.e. the number is
$$
2(n-1) + (n-2) + (n-3) + ........... + 3 + 2 + 1
=\frac{1}{2}(n+1)n - 1
\eqno{(3)}
$$
In the product basis total number of operators will be twice the number
calculated above. So the number is 
$$
(n+2)(n-1).
\eqno{(4)}
$$

In the case of coupled basis, we will have $(n-1)$ casimir operators for 
each of the two $SU(n)$ groups. Also we will get $(n-1)$ number of operators 
analogous to $J^2$ in case of $SU(2)$ group (in that case the number was
$2-1 = 1$). Then there will be $(n-1)$ operators analogous to $J_z$ in the
case of $SU(2)$. Finally if we consider the casimir operators of each lower 
group the total number of operators will be 
$$
(n-2) + (n-3) + ........... + 3 +2 +1 = \frac{(n-1)(n-2)}{2}
\eqno{(5)}
$$
So the total number of operators in this case equals to
$$
2(n-1) + (n-1) + (n-1) + \frac{(n-1)(n-2)}{2} = \frac{1}{2}(n^2 + 5n - 6)
\eqno{(6)}
$$

Now the difference in the number of operators between the product basis
and the coupled basis is (from expressions (4) and (6))
$$
(n+2)(n-1) - \frac{1}{2}(n^2 + 5n - 6) = \frac{1}{2}(n-1)(n-2).
\eqno{(7)}
$$
Clearly the above number gives correct result for $SU(2)$ and $SU(3)$ which
we have already discussed.

Now if we consider the next higher group which is $SU(4)$, in the product
basis the complete set of operators is 

$A^3(1), A^3(2), B^3(1), B^3(2), C^3(1), C^3(2), G^3(1), G^3(2),
F^2(1), F^2(2), I^2(1), I^2(2), I_3(1), I_3(2), Y(1), Y(2),
Z(1), Z(2)$ 

and the total number is $18$. But in the coupled basis 
the total number of commuting operators is only $15$ which are given below,

$A^3, A^3(1), A^3(2), B^3, B^3(1), B^3(2), C^3, C^3(1), C^3(2),
G^3, F^2, I^2, I_3, Y, Z$.

Hence the difference in the number of operators is $3$, which we can also verify
from our general expression (7) by putting $n=4$.

Similarly we can verify for the cases of higher dimensional group. Thus,
if we know  the number of complete set of operator 
in a particular representation then we can calculate number of non
complete set of operator in that representation and vice-versa.
Here we have introduced a {\it general rule}, which indicates
number difference between complete and non-complete set of
operatiors for any particular representation.  

\vskip1cm

It is a great pleasure for us to thank Prof. Amitabha Raychaudhuri for 
many helpful discussions.


\begin{references}

\bibitem {} J.J. De Swart,  Review of Modern Physics, {\bf 35}, no.-4  (1963).

\bibitem {} G. B. Lichtenberg, (Second Ed.) in {\it Unitary Symmetry and 
elementary Particles} (New York, San Francisco, London: Academic Press, 1978).

\bibitem {} H. Georgi, in {\it Lie Algebras in Particle Physics} 
(London, Amsterdam, Don Mills, Ontario, Sydney: The Benjamin/Cummings
Publishing Company, Inc., 1982).

\end{references}
\end{document}